\documentclass[useAMS,usenatbib,usegraphicx]{mn2e}
\usepackage{graphicx}

\title[New visual companions]
{New visual companions of solar-type stars within 25\,pc}
\author[R. Chini et al.]
{R. Chini$^{1,2}$, K. Fuhrmann$^{1,3}$, A. Barr$^1$, F. Pozo$^1$, C. Westhues$^1$, and K. Hodapp$^4$\\
$^{1}$Astronomisches Institut, Ruhr-Universit\"at Bochum, Universit\"atsstra\ss{}e 150, D-44801 Bochum, Germany\\
$^{2}$Instituto de Astronom\'{\i}a, Universidad Cat\'{o}lica del Norte, Avenida Angamos 0610, Casilla 1280, Antofagasta, Chile\\
$^{3}$Isaac Newton Group of Telescopes, Apartado 321, E-38700 Santa Cruz de La Palma, Spain\\
$^4$Institute for Astronomy, University of Hawaii, 640 N. Aohoku Place, Hilo, HI 96720, USA}
\begin{document}

\date{}

\pagerange{\pageref{firstpage}--\pageref{lastpage}} \pubyear{2012}
\maketitle

\label{firstpage}

\begin{abstract}
We report the discovery of faint common-proper-motion companions to the nearby southern solar-type stars HD\,43162, HD\,67199, HD\,114837, HD\,114853, HD\,129502, HD\,165185, HD\,197214, and HD\,212330 from near-infrared imaging and astrometry. We also confirm the previously identified tertiary components around HD\,165401 and HD\,188088. Since the majority of these stars were already known as binaries, they ascend now to higher-level systems. A particularly interesting case is the G6.5\,V BY~Dra-type variable HD\,43162, which harbors two common-proper-motion companions at distances of 410\,AU and 2740\,AU. Our limited study shows that the inventory of common-proper-motion companions around nearby bright stars is still not completely known.
\end{abstract}


\section{Introduction}

In recent years, a large amount of work has been invested on sources of the Solar Neighborhood. A main driver is certainly the study of extra-solar planets or planetary systems and their host stars. Since it is also realized that a considerable fraction of the latter are non-single, the question arises to what extent the existing stellar companions may affect the overall planet formation. Likewise, undiscovered companions influence the calculated properties of the host star and as such the derived properties of the extra-solar planet. As a result, there is a vital search for as yet unknown companions, mostly white, red, and brown dwarfs. However -- and more generally speaking -- it is also appreciated that it is of principal importance to know the local inventory as this has far-reaching consequences on all scales of astrophysical research. The locally existing degenerates mostly contribute to issues of the mass density, the stellar populations, and the early Galactic epochs.

Simultaneously, there is a major interest in the star formation process in general because evidence is growing that stars often are created as multiple systems \citep{GETAL07}. On the other hand, the formation of very wide binaries is difficult to understand, because their observed separation can exceed the typical size of a collapsing cloud core. Recent observations have shown that very wide binaries are frequently members of triple systems \citep{Faherty10,Law10} and that close binaries often have a distant third companion \citep{Allen12,Tokovinin06,Tokovinin02}.

Dynamical interactions between members of such systems lead to close encounters which generally lead to the ejection of the lowest mass member and thus to the formation of a stable binary \citep{Delgado04}. However, a bound triple with long-term stability can also be created in the presence of a gravitational potential \citep{Reipurth10} which is frequently fulfilled when the newborn stars are still embedded in their cloud cores \citep{Reipurth00}. Recently, \citet{Reipurth12} presented $N-$body simulations of the dynamical evolution of newborn triple systems that match observations of very wide systems  \citep{Chaname04,Lepine07,TL12}. The simulations show that although triples start as very compact systems, one component may be dynamically scattered into a very distant orbit on timescales of millions of years. Simultaneously, the orbits of the other two stars shrink, resulting in a very close binary system. The study of the nearby field stars, as the final product of these initial dynamical processes, serves as an important observational constraint in this context.

Advances in high-precision radial velocities (\emph{RV}s) and high-angular resolution imaging have led to the discovery of numerous new neighbors in the solar vicinity. The latter observations also benefit from inclusion of as yet less explored wavelength regions -- such as the infrared -- where cool sources are better traceable. Yet, highly-resolved imaging observations are often confined to the innermost few arcseconds around a primary target, i.e., they do not reach beyond the classical regions in the search for visual binaries.

\begin{figure*}
\resizebox{145mm}{!}
{\includegraphics[bbllx=72pt,bblly=360pt,bburx=520pt,bbury=490pt]{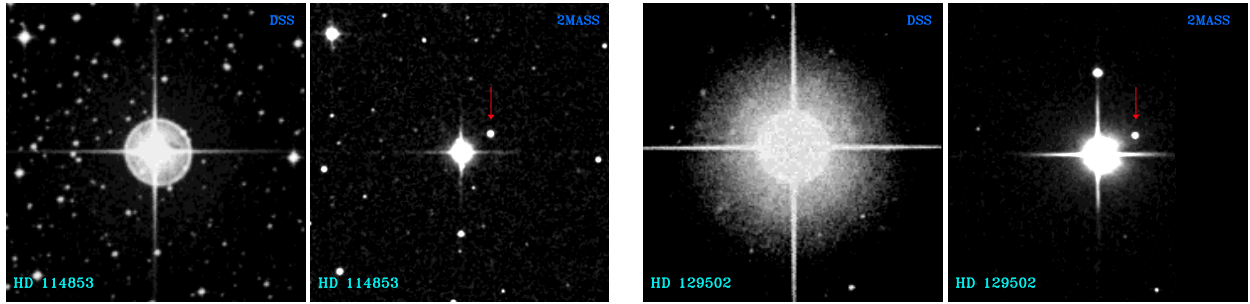}}
\caption{DSS and 2MASS $K_s$ images of HD\,114853 ($V=6.9$\,mag, left) and $\mu$\,Vir~=~HD\,129502 ($V=3.9$\,mag, right); the field size is $5\times5$\,arcmin, north is up and east is to the left. The red arrows mark the common-proper-motion companions found in our investigation. These are almost invisible in the optical, but clearly present in the infrared images.}
\label{FIG_HD114853_DSS}
\end{figure*}

Large multi-epoch star surveys, on the other hand, are mostly concerned with the search for common-proper-motion (CPM) objects that can even surpass spatial distances of more than one degree. Yet, this approach often refers to photographic plate material for the first-epoch observations and this in turn leads to the well-known peculiarity that a considerable region around luminous and mostly nearby stars generally remains inaccessible. Depending on how bright the sources actually are, this may reach from a few arcseconds and can easily exceed one arcminute for stars of third magnitude or brighter. Two examples where close companions may be easily missed are displayed in Fig.~\ref{FIG_HD114853_DSS}. ESO Digitized Sky Survey plates with a $5\times5$\,arcmin field are compared to 2MASS infrared $K_s$ images for HD\,114853 (left) and $\mu$\,Vir = HD\,129502 (right). In the case of HD\,114853 ($V = 6.9$\,mag) the companion is hidden behind a diffraction pattern of component A, in the case of $\mu$\,Vir = HD\,129502 ($V = 3.9$\,mag) the companion is masked by the bright straylight of the primary.

In consequence, this means that one usually encounters many unexplored ``white patches" on celestial maps, depending on which survey one intends to rely on. The situation is imaginably -- and ironically -- most serious for the nearby stars. For a particular example, we mention the prominent fourth-magnitude exoplanet host star $\upsilon$\,And, whose $\rho \simeq 55$\,arcsec distant M\,dwarf companion was only uncovered a decade ago by \citet{LETAL02}. From this $\upsilon$\,And~B discovery paper and as also more recently discussed in \citet[][their Fig.~7]{Raghavan10}, it is most obvious that a secondary at about half the spatial distance of $\upsilon$\,And~B would be difficult to discern in the light of its primary.

Thus, and as the most recent observations by \citet{Allen12} again demonstrate, our census of the nearby CPM companions is far from complete. The situation is particularly severe in the Southern Hemisphere, where -- traditionally -- there has been a less rigorous monitoring of celestial objects. In this work, we present near-infrared imaging and astrometry on solar neighborhood stars within 25\,pc. In particular, we report the discovery of seven faint CPM companions to solar-type primaries at spatial distances ranging from 12\,arcsec to 42\,arcsec, i.e., in regions where previous photographic material mostly fails.

\section{Data}

\begin{figure}
\centering
	\includegraphics[width=\columnwidth]{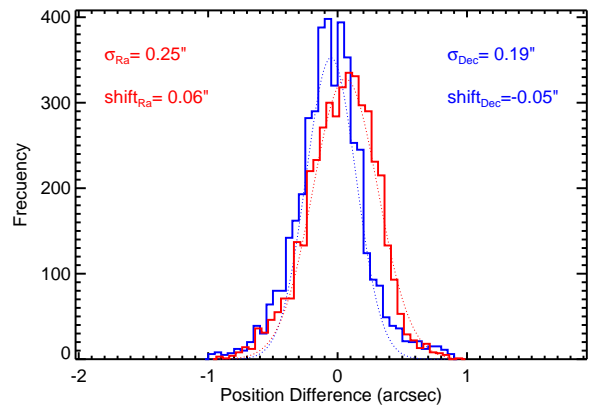}

\caption{Astrometric accuracy. The histogram shows the distribution of the $RA$ (red) and $Dec$ (blue) offsets obtained by matching the coordinates (within 1 arcsec) of all stars detected in the IRIS frames with the 2MASS catalogue. The width of the respective gaussian distribution displays the scatter $\sigma_{RA} = 0.25\arcsec$ and $\sigma_{Dec} = 0.19\arcsec$. We attribute the small systematic deviations in $RA$ ($0.06\arcsec$) and $Dec$ ($-0.05\arcsec$) to the SCAMP astrometry process in the data reduction.}
\label{FIG_hist.dcf.cpm}
\end{figure}

For our first-epoch observations we make use of the Two Micron All Sky Survey (2MASS) point source catalog \citep{SETAL06}, whereas for the second and third epoch we refer to the recent Wide-field Infrared Survey Explorer (WISE) data \citep{WETAL10} and our own measurements. The latter were taken with the IRIS telescope of the Universit\"atssternwarte Bochum near Cerro Armazones in Chile \citep{HETAL10} during October and November 2012. We employed a dithering pattern with 20 exposures on the object and 20 exposures on a nearby sky field with an integration time of 10\,s each. Our IRAF-based reduction pipeline cares about the common corrections like flat, sky, bad pixel, etc. and compensates the geometrical image distortion with the IRAF tool \emph{geotran}; for that purpose we use a theoretically calculated Zemax matrix created for our IRIS camera. Next we construct catalogs with $xy-$coordinates for all sources on the individual frames (\emph{SExtractor}); simultaneously, we download the corresponding coordinates from the 2MASS catalog. Both coordinate lists -- IRIS and 2MASS --  are matched (\emph{scamp}) and the resulting astrometry is stored in the Fits-header of each individual exposure; frames with non-unique astrometric solutions were rejected. The useful individual frames were then shifted against each other (\emph{wregister}) and combined into a single frame (\emph{imcombine}) that contains the astrometry. Eventually, the tool \emph{SExtractor} provides the final source list from the combined image with \emph{WCS} positions. In all cases CPM pairs were easily discovered by eye by just blinking the IRIS and 2MASS images and compare their displacement with respect to the numerous field stars on the frames. The astrometric accuracy of the IRIS data is displayed in Fig.~\ref{FIG_hist.dcf.cpm}.

\begin{table*}
  \caption{Common proper motion candidates of nearby solar-type stars.}
   \label{Photometry}
     \begin{center}
         \begin{tabular}{llrrrrrrrrrc}
             \hline\hline
             \multicolumn{1}{c}{Name}
           & \multicolumn{1}{c}{2MASS}
           & \multicolumn{1}{c}{RA}
           & \multicolumn{1}{c}{Dec}
           & \multicolumn{1}{c}{$B$}
           & \multicolumn{1}{c}{$V$}
           & \multicolumn{1}{c}{$R$}
           & \multicolumn{1}{c}{$I$}
           & \multicolumn{1}{c}{$J$}
           & \multicolumn{1}{c}{$H$}
           & \multicolumn{1}{c}{$K$}
           & \multicolumn{1}{c}{Spectral}\\
             \multicolumn{1}{c}{}
           & \multicolumn{1}{c}{}
           & \multicolumn{1}{c}{J2000}
           & \multicolumn{1}{c}{J2000}
           & \multicolumn{1}{c}{mag}
           & \multicolumn{1}{c}{mag}
           & \multicolumn{1}{c}{mag}
           & \multicolumn{1}{c}{mag}
           & \multicolumn{1}{c}{mag}
           & \multicolumn{1}{c}{mag}
           & \multicolumn{1}{c}{mag}
           & \multicolumn{1}{c}{Type}\\
             \hline
HD\,43162     & J06134528$-$2351433 &  93.438699 & -23.862036 &  7.04 &  6.37 &  5.94 &  5.61 &  5.13 &  4.86 &  4.73 & G6.5\,V \\
HD\,43162\,B  & J06134539$-$2352077 &  93.439129 & -23.868814 & 12.46 &       &       &       &  8.37 &  7.79 &  7.53 & M3.5    \\
HD\,43162\,C  & J06134717$-$2354250 &  93.446545 & -23.906952 & 14.32 & 12.96 & 11.78 & 10.21 &  8.72 &  8.16 &  7.87 & M4      \\
\noalign{\smallskip}
HD\,67199     & J08023117$-$6601153 & 120.629906 & -66.020927 &  8.05 &  7.17 &  6.69 &  6.26 &  5.65 &  5.23 &  5.12 & K2\,V   \\
HD\,67199\,B  & J08023164$-$6601332 & 120.631841 & -66.025909 &       & 17.00 &       &       &  9.07 &  8.47 &  8.22 &         \\
\noalign{\smallskip}
HD\,114837    & J13141513$-$5906114 & 198.563072 & -59.103184 &  5.40 &  4.92 &  5.46 &  8.83 &  3.94 &  3.57 &  3.61 & F6\,V   \\
HD\,114837\,B &                     & 198.563750 & -59.102263 &       & 10.2: &       &       &       &       &       &         \\
\noalign{\smallskip}
HD\,114853    & J13135221$-$4511090 & 198.467550 & -45.185856 &  7.58 &  6.94 &  6.56 &  6.23 &  5.73 &  5.43 &  5.32 & G1.5\,V \\
HD\,114853\,B & J13134938$-$4510501 & 198.455777 & -45.180599 &       &       &       &       & 11.12 & 10.58 & 10.31 &         \\
\noalign{\smallskip}
HD\,129502    & J14430363$-$0539291 & 220.765132 &  -5.658098 &  4.36 &  3.90 &  3.62 &  3.62 &  3.34 &  3.07 &  3.04 & F2\,V   \\
HD\,129502\,B & J14430110$-$0539099 & 220.754618 &  -5.652752 &       &       &       &       & 10.72 & 10.17 &  9.88 &         \\
\noalign{\smallskip}
HD\,165185    & J18062370$-$3601113 & 271.598755 & -36.019825 &  6.52 &  5.95 &  5.58 &       &  4.84 &  4.61 &  4.47 & G0\,V   \\
HD\,165185\,B & J18062369$-$3601237 & 271.598738 & -36.023262 & 12.36 &       &       &       &  8.95 &  8.41 &  8.11 & M0      \\
\noalign{\smallskip}
HD\,165401    & J18053746$+$0439255 & 271.406089 &  4.657102 &  7.40 &  6.79 &  6.43 & 6.09  &  5.66 &  5.35 &  5.25 & G0\,V   \\
HD\,165401\,B & J18053758$+$0439405 & 271.406591 &  4.661277 &       &       &       &       & 10.26 &  9.93 &  9.63 &         \\
\noalign{\smallskip}
HD\,188088    & J19541775$-$2356275 & 298.573991 & -23.940987 &  7.20 &  6.18 &  5.58 &  5.09 &  4.75 &  4.16 &  4.04 & K2-3\,V \\
HD\,188088\,B & J19542064$-$2356398 & 298.586008 & -23.944410 &       &       & 15.11 & 11.94 &  9.79 &  9.19 &  8.84 & M5      \\
\noalign{\smallskip}
HD\,197214    & J20431599$-$2925260 & 310.816662 & -29.423893 &  7.64 &  6.96 &  6.58 &  6.23 &  5.71 &  5.38 &  5.31 & G6\,V   \\
HD\,197214\,B & J20431469$-$2925217 & 310.811224 & -29.422695 &       &       &       &       & 11.42 & 10.87 & 10.57 &      \\
\noalign{\smallskip}
HD\,212330    & J22245638$-$5747508 & 336.234920 & -57.797470 &  5.99 &  5.31 &  4.90 &  4.58 &  4.32 &  3.89 &  3.85 & G2\,IV-V\\
HD\,212330\,B & J22245358$-$5748000 & 336.223277 & -57.800026 &       &       &       &       &  9.61 &  9.06 &  8.77 &         \\
\hline
\end{tabular}

\end{center}
\footnotesize{The optical photometry was taken from SIMBAD, $JHK$ from 2MASS. $VRI$ data for HD\,43162\,C are from \citet{Raghavan10}. \hfill}
\end{table*}

\begin{figure}
	\includegraphics*[width=84mm]{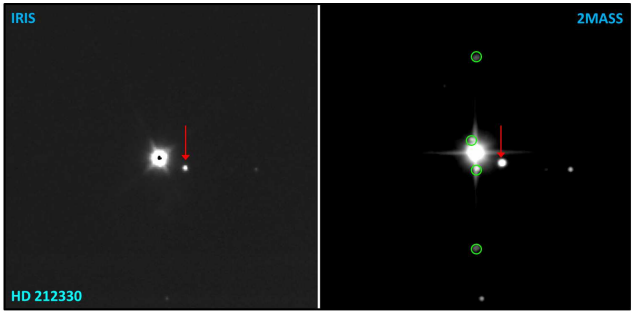}

\caption{$K_s$ images of HD\,212330 from IRIS (left) and 2MASS (right); the field size is $4.5 \times 4.5 $\,arcmin, north is up and east is to the left. The red arrow marks the corresponding companion that has moved toward the south in the IRIS image. Numerous ghosts on the 2MASS image are encircled.}
\label{FIG_HD212330}
\end{figure}

In many cases, the 2MASS frames of our bright targets display "ghosts" which are generally oriented in north-south direction and typically occur in a symmetric pattern with respect to the true stellar image. Fig.~\ref{FIG_HD212330} shows an example for such ghosts around HD\,212330 which become obvious when comparing the IRIS and the 2MASS image. The IRIS frame also shows a clear displacement of the CPM pair to the south with respect to other stars in the field.

\section{The sample}

We are currently investigating the multiplicity of nearby stars ($d \le 25$\,pc) with absolute magnitudes $M_V \le 6.0$, i.e. stars with spectral types earlier than about K2\,V. To search for CPM candidates we have checked the 2MASS catalog for potential companions within 1\arcmin among our sample of $\sim 270$ stars accessible from the southern hemisphere. Among 44 promising targets we found 10 objects with common proper motion candidates. In the following we introduce each individual primary and describe the properties of the CPM candidates. The distances listed for the primaries are from the revised Hipparcos analysis \citep{vLeeuwen07}; their $V-$magnitudes were obtained from the SIMBAD service.

\subsection{New Candidate Companions}

\textbf{HD\,43162} (HIP\,29568, V352\,CMa) is a BY~Dra variable of spectral type G6.5\,V \citep{Gray06} with $V = 6.37$ at a distance of 17\,pc. The nine \emph{RV} measurements by \citet{Abt06} do not show any significant variation ($v_r = +21.91 \pm 0.09$\,km~s$^{-1}$); this \emph{RV} is consistent with the entry by \citet{Nordstrom04}. The \emph{Hipparcos} catalog lists a Hvar Type `U' suggesting that its photometric variability might be due to an inner Aa-Ab binary. \citet{Raghavan10} report the discovery of a CPM companion at a distance of 164\arcsec (2740\,AU). Our data corroborate this discovery but show an additional closer CPM companion at 24\arcsec (410\,AU), turning HD\,43162 into a triple system. The spectral type and distance of component B is M3.5 and 20\,pc in \citet{Riaz06}. Component C is slightly fainter and its photometry suggests that it is of spectral class M4 \citep{Raghavan10}.\\

\noindent
\textbf{HD\,67199} (HIP\,39342) is a star of spectral type K2\,V with $V = 7.17$ at a distance of 17\,pc. \citet{Raghavan10} report an inner Ab companion candidate with a 3.1\,$\sigma$ significance. The distant B companion from our work (cf. Fig.~\ref{FIG_CPMS_1}) lies 18\arcsec (310\,AU) toward southeast and is about 3 mag fainter at $JHK$; its colors suggest an M-type.\\

\begin{figure}
	\includegraphics*[width=84mm]{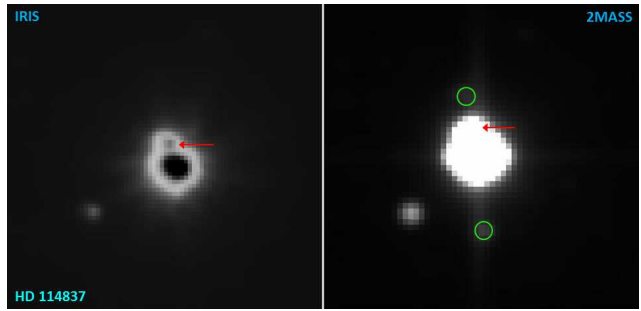}

\caption{$K_s$ images of HD\,114837 from IRIS (left) and 2MASS (right); the field size is $55 \times 55 $\,arcsec, north is up and east is to the left. The red arrow marks the corresponding companion -- barely to be seen in the 2MASS frame -- that has moved toward the southwest in the IRIS image. Two ghosts on the 2MASS image are encircled.}
\label{FIG_HD114837}
\end{figure}

\begin{figure*}
\resizebox{230mm}{!}{\includegraphics[bbllx=80pt,bblly=380pt,bburx=520pt,bbury=710pt]{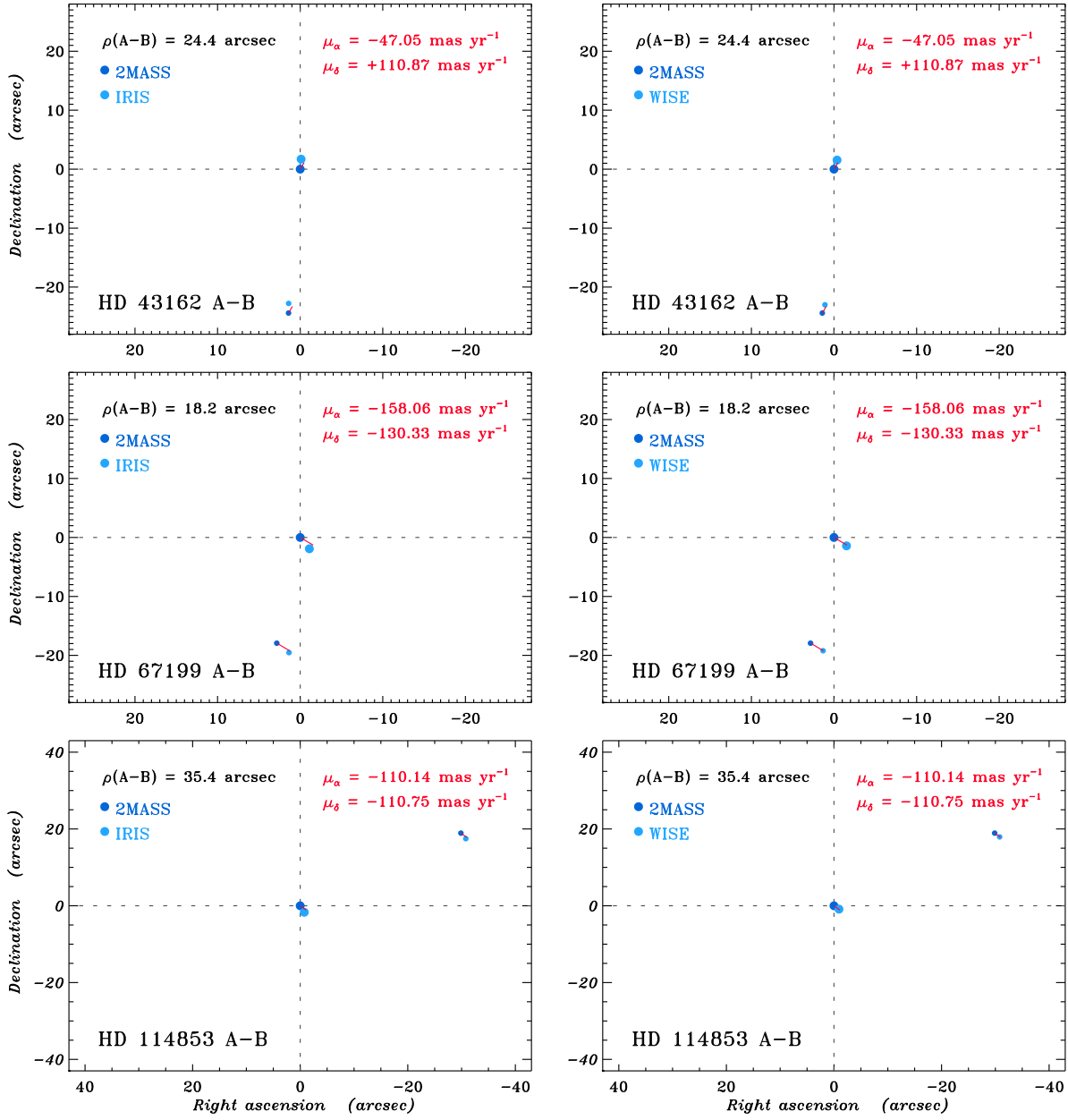}}
\caption{Relative stellar positions of the CPM pairs HD\,43162~A-B, HD\,67199~A-B, and HD\,114853~A-B (top to bottom), with 2MASS as the first-epoch observations and with reference to IRIS (left column) and WISE (right column) astrometric data for a second and third epoch. In all panels the primary position at the 2MASS epoch defines the centre of the coordinate system. Red numbers refer to the {\em Hipparcos} proper motion and the red lines denote a 10-yr {\em Hipparcos}-based trajectory. Note that the time span for the IRIS observations is slightly larger compared to WISE.}
\label{FIG_CPMS_1}
\end{figure*}

\noindent
\textbf{HD\,114837} (HIP\,64583) is an F6\,V star with $V = 4.92$ at a distance of 18\,pc. The Washington Double Star Catalog (WDS) mentions a single observation in 1928 by Finsen with a 5.3\,mag fainter visual companion at $\rho = 2.7\arcsec$. The 2MASS image shows a slightly elongated stellar image suggesting that it might be indeed a binary. Our IRIS image clearly resolves the two stars and confirms the CPM of the system (Fig.~\ref{FIG_HD114837}). The separation of 4.2\arcsec explains why the companion could not be detected by \citet{Tokovinin12} who observed with a frame size of only $\sim 6\arcsec \times 6\arcsec$. The companion is about 5 mag fainter at $K$, its proximity to component A, however, does not allow a precise photometry.\\

\noindent
\textbf{HD\,114853} (HIP\,64550) is a solar-like G1.5\,V star with $V = 6.94$ at a distance of 24\,pc. As one can deduce from Fig.~\ref{FIG_CPMS_1}, it possesses a rather small proper motion. Its considerable \emph{RV} $v_{r,A} \simeq +58\,$km~s$^{-1}$, however, is a good criterion for follow-up high-resolution spectroscopy of the distant companion (860\,AU). This companion, displayed in Fig.~\ref{FIG_HD114853_DSS}, is about 5\,mag fainter at $JHK$. \\

\noindent
\textbf{HD\,129502} (HIP\,71957, $\mu$\,Vir) is a $V = 3.90$ bright F2\,V star at a distance of 18\,pc. As already mentioned in \citet{FC12} the primary is itself a likely close binary, as implied from interferometric VLTI observations by \citet{RPD09}. The distant B companion at a linear separation of 770\,AU in Fig.~\ref{FIG_HD114853_DSS} is 7\,mag fainter at $JHK$.  \\

\noindent
\textbf{HD\,165185} (HIP\,88694) is a G0\,V star with $V = 5.95$ at a distance of 18\,pc. For our 12\arcsec distant CPM candidate \citet{Riaz06} give a rough distance of 51\,pc and classify it spectroscopically as M0; it is 4\,mag fainter at $JHK$ than component A. This apparent controversy in the distances can be resolved by considering that the estimate for the CPM candidate was obtained from its TiO5 index which may involve large errors. As stated by \citet{Riaz06} the average uncertainties in such distance estimates are of the order of $\pm\, 37\%$. We therefore suggest that our CPM candidate is a physical companion of HD\,165185.\\

\noindent
\textbf{HD\,197214} (HIP\,102264) is a star of spectral type G6\,V with $V = 6.96$ at a distance of 22\,pc. The large velocity scatter $\sigma = 4.1$\,km~s$^{-1}$ from only four measurements around a mean \emph{RV} of $-21.88\,$km~s$^{-1}$ \citep{Nidever02} suggests it to be a spectroscopic binary. \citet{Nordstrom04} give a zero probability that the observed scatter of the \emph{RV}s is due to random observational errors only. Likewise the \emph{RV} measurements by \citet{Abt06} also show substantial variations corroborating the classification as an SB1. The error in the \emph{Hipparcos} parallax ($\pi = 44.83 \pm 0.91$\,mas) is rather high, indicating an astrometric binary with a period of the order of years.

\citet{Riaz06} classify the CPM candidate as a K5 star at a distance of 232\,pc; this distance estimate is again based on the TiO5 index and as such on the $J-$band magnitude. However, the $JHK$ magnitudes quoted by \citet{Riaz06} differ slightly from those in the 2MASS catalog. Nevertheless, both the photometry listed in Table 1 of \citet{Riaz06} and the 2MASS photometry given in Table~\ref{Photometry} are not compatible with spectral type K5. The intrinsic colors of a K5 type are $J - H = 0.42$ and $H - K = 0.20$ \citep{Ducati01} while the observed 2MASS colors are $J - H = 0.55$ and $H - K = 0.30$, i.e. they rather are in line with spectral class M.

Additionally, there is a kinematic argument: The $UVW$ velocities of HD\,197214 are $U,V,W = 0,-21,18$\,km/s, i.e. a normal disk star with $V_{\rm rot} \sim 200$\,km/s (assuming $V_{\rm rot} = V + 220$\,km/s). If component B were located at a distance of 232\,pc its $U,V,W$ velocities were $50,-226,11$\,km/s, i.e. the star would be a halo object with no galactic rotation ($V_{\rm rot} \sim 0$\,km/s). It is implausible to assume that HD\,192714 shares its velocity with a halo object for many years. Given that the distance of component B is about 390\,AU crude estimates show that its influence on the velocity scatter of the primary is of the order of only m/s/yr. We therefore conclude that the observed velocity scatter of $\sigma = 4.1$\,km~s$^{-1}$ \citep{Nidever02} must be due to a close, invisible component Ab while the CPM candidate found in our study is a third member of the system.\\

\noindent
\textbf{HD\,212330} (HIP\,110649) is an astrometric binary of spectral type G2\,IV-V with $V = 5.31$ at a distance of 21\,pc; it is also a single-line spectroscopic binary with $\Delta m < 5$\,mag \citep{Nordstrom04}. \citet{Tokovinin10} observed this star with speckle interferometry but did not find any visible component. This suggests that component Ab is significantly closer than 1 arcsec to its primary. The CPM candidate found in the present study has a projected distance of $\rho = 24\farcs{2}$ which qualifies it as a new, third member of the system. A model atmosphere analysis and a brief discussion of HD\,212330 was already given in \citet{FC12}. The distant B component is about 5\,mag fainter at $JHK$.  \\

\subsection{Previously Identified Candidate Companions}

\noindent
\textbf{HD\,165401} (HIP\,88622) is an old Population\,II star of spectral type G0\,V \citep{Gray03} with $V = 6.79$ at a distance of 24\,pc. \citet{Nordstrom04} list 19 \emph{RV} measurements over 4451 days which show a variation of 0.9\,km~s$^{-1}$. These authors assign a ``zero probability of constant $R_{\rm vel}$" suggesting that it is a spectroscopic binary. The comparison with its $\rho \simeq 15\arcsec$ wide companion shows that it possesses the same extreme \emph{RV} $v_r \simeq -120\,$km~s$^{-1}$ as the primary \citep{F08}. The comparison of the 2MASS, WISE, and IRIS astrometric data in Fig.~\ref{FIG_CPMS_3} now confirms this result in terms of the proper motions. This companion is about 4.5\,mag fainter at $JHK$.  \\

\noindent
\textbf{HD\,188088} (HIP\,97944, V4200~Sgr) is a BY~Dra variable of spectral type K2-3\,V with $V = 6.18$ at a distance of 14\,pc. It is a near-equal-mass double-lined spectroscopic binary with an orbital period $P=46.8$\,d, an eccentricity $e=0.69$, and minimum masses for the components of 0.85\,$M_\odot$ \citep{Fekel83}. Recently \citet{Allen12} reported a tertiary companion candidate which displays consistent CPM over a baseline of 50 years. It is about 5 mag fainter at $JHK$; its spectral type was determined as M5. The WISE and IRIS data confirm the common-proper-motion status.

\section{Discussion}

Except for the controversial B component of HD\,197214, all spectroscopically classified CPM candidates of this work turn out to be M dwarfs. Their mean $J-H$ and $H-K$ colours are 0.54 and 0.30, respectively. The photometry of the companions without spectra yields mean values of $J-H = 0.51$ and $H-K = 0.28$. Therefore we conclude that the unclassified companions are also low-mass objects in the K to M range. Likewise, the brightness difference between components A and B corroborate such spectral classes.

\begin{table}
 \caption{Possible triple or higher level systems.}
   \label{triple}

         \begin{tabular}{lcc}
            \hline\hline
Name       & Distant CPM & Primary \\
           &     [AU]    &         \\
            \hline
HD 43162  &   410  & BY Dra-type variable\\
          &  2740  & Hvar Type `U' $^1$ \\

HD 67199  &   310  & Aa-Ab candidate $^2$\\

HD 114853  &   860  & \\

HD 129502  &   770  & Aa-Ab candidate $^3$\\

HD 165185  &   220  & Hvar Type `U' $^1$ \\

HD 165401  &   360  & SB1 \\

HD 188088  &   580  & SB2 \\

HD 197214  &   390  & SB1 \\

HD 212330  &   500  & SB1 \\

\hline
\end{tabular}\\
\footnotesize{
Notes:\\
$^1$ might indicate Aa-Ab subsystem \\
$^2$ $3.1 \sigma$ proper motion difference between Hipparcos and Tycho\,2 \\
$^3$ interferometric diameter observations\\
HD\,114837 is here excluded, since its B component has a linear separation of only a few arcsec corresponding to less than 50\,AU when W.S. Finsen first observed it in 1928.
}
\end{table}

Almost each of the nine CPM systems may contain an inner Aa-Ab binary as demonstrated by the supplementary information in Table~\ref{triple}. This boosts the majority of these stars toward triple or higher level systems.

Table~\ref{statistic} summarizes a preliminary CPM statistics for bright Population I stars within 25\,pc: Our ``northern sample" contains 274 stars north of Dec\,$= -15^\circ$ \citep{F11}, the ``southern sample" comprises 155 stars \citep{FC12}. The spectral types cover a range from B to K2; however, there is just one B star (Regulus) and only about 25 A stars which means that more than 90\% of the sample are solar-type stars of type F, G and early K. We have arbitrarily divided our samples in two groups with projected orbital distances larger than 100 and 500\,AU, respectively. The present paper adds one "distant" CPM in the north (HD\,129502) and five CPMs in the south two of which have orbital distances $> 500$\,AU. The comparison between the CPM percentage in the north and south suggests that one can expect about 12 further southern CPMs of which 10 have orbital distances larger than 500\,AU. A possible future detection of those stars, however, will not mean that all CPMs of the nearby bright stars are known.

\begin{table}
 \caption{CPM statistics.}
   \label{statistic}

         \begin{tabular}{lccc}
            \hline\hline
             \multicolumn{1}{l}{All}
           & \multicolumn{1}{c}{CPMs $> 100$\,AU}
           & \multicolumn{1}{c}{CPMs $> 500$\,AU}
           & \multicolumn{1}{c}{}\\
\hline
             \multicolumn{4}{c}{Northern sample}\\
\hline
274        &  75 (27.4\%)       &  47 (17.2\%)        & prev.  \\
274        &  76 (27.7\%)       &  48 (17.5\%)        & now    \\
\hline
             \multicolumn{4}{c}{Southern sample}\\
\hline
155        &  26 (16.8\%)       &  15 ( 9.7\%)        & prev.  \\
155        &  31 (20.0\%)       &  17 (11.0\%)        & now    \\
           &        12          &        10           & "missing" \\
\hline
\end{tabular}
\end{table}

\section*{Acknowledgments}

This research has made use of the SIMBAD database, operated at CDS, Strasbourg, France. This publication makes use of data products from the Two Micron All Sky Survey, which is a joint project of the University of Massachusetts and the Infrared Processing and Analysis Center/California Institute of Technology, funded by the National Aeronautics and Space Administration and the National Science Foundation. This publication makes use of data products from the Wide-field Infrared Survey Explorer, which is a joint project of the University of California, Los Angeles, and the Jet Propulsion Laboratory/California Institute of Technology, also funded by NASA. This publication is supported as a project of the Nordrhein-Westf\"alische Akademie der Wissenschaften und der K\"unste in the framework of the academy program by the Federal Republic of Germany and the state Nordrhein-Westfalen. K.F. acknowledges support from the DFG grant FU\,198/10-1 and productive conversations with Daniel, Fabian, and Maria Muhs.

\begin{figure*}
\resizebox{230mm}{!}{\includegraphics[bbllx=95pt,bblly=380pt,bburx=520pt,bbury=710pt]{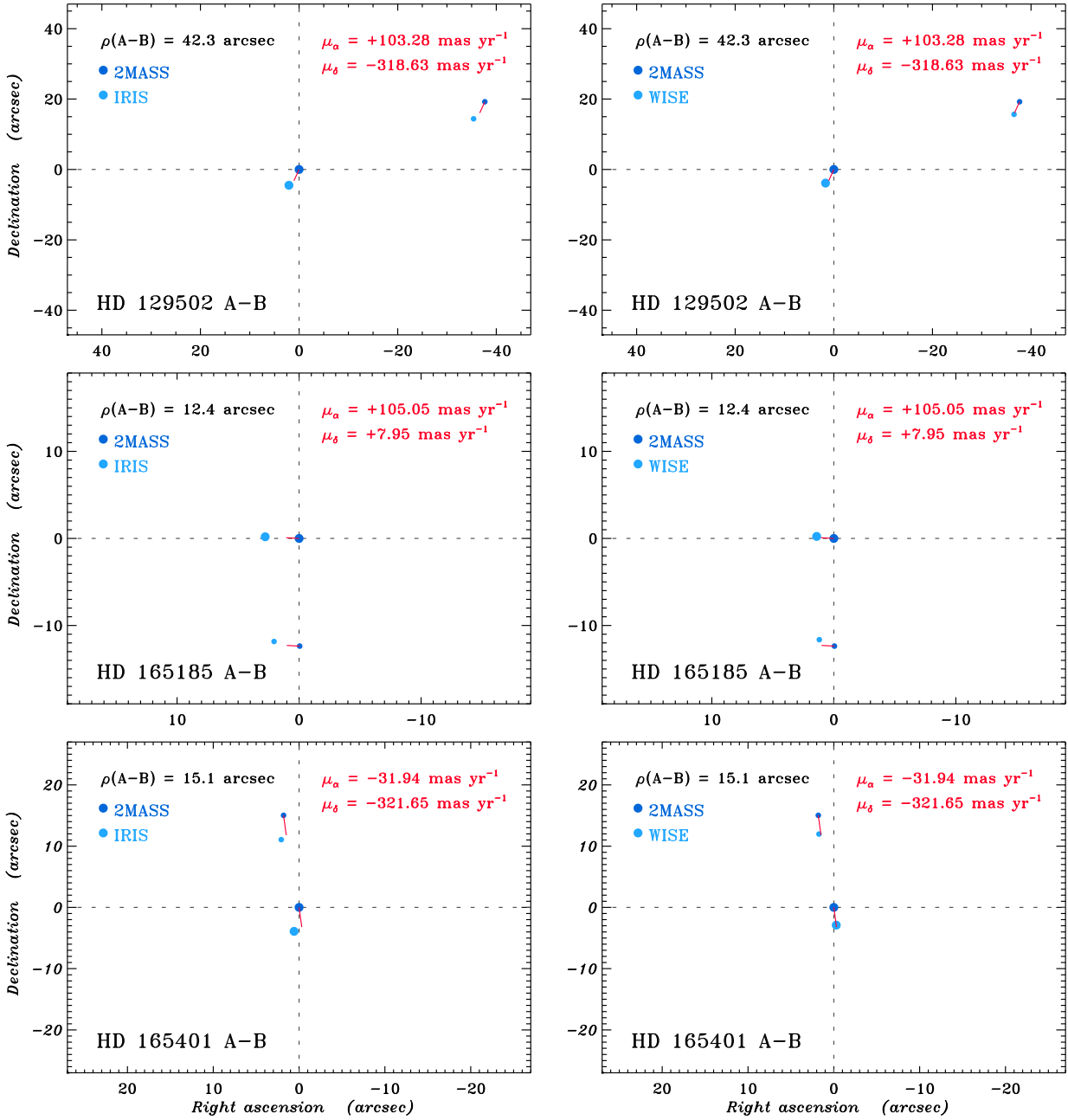}}
\caption{Same as Fig.~\ref{FIG_CPMS_1}, but for HD\,129502~A-B, HD\,165185~A-B, and HD\,165401~A-B.}
\label{FIG_CPMS_2}
\end{figure*}

\begin{figure*}
\resizebox{230mm}{!}{\includegraphics[bbllx=80pt,bblly=380pt,bburx=520pt,bbury=710pt]{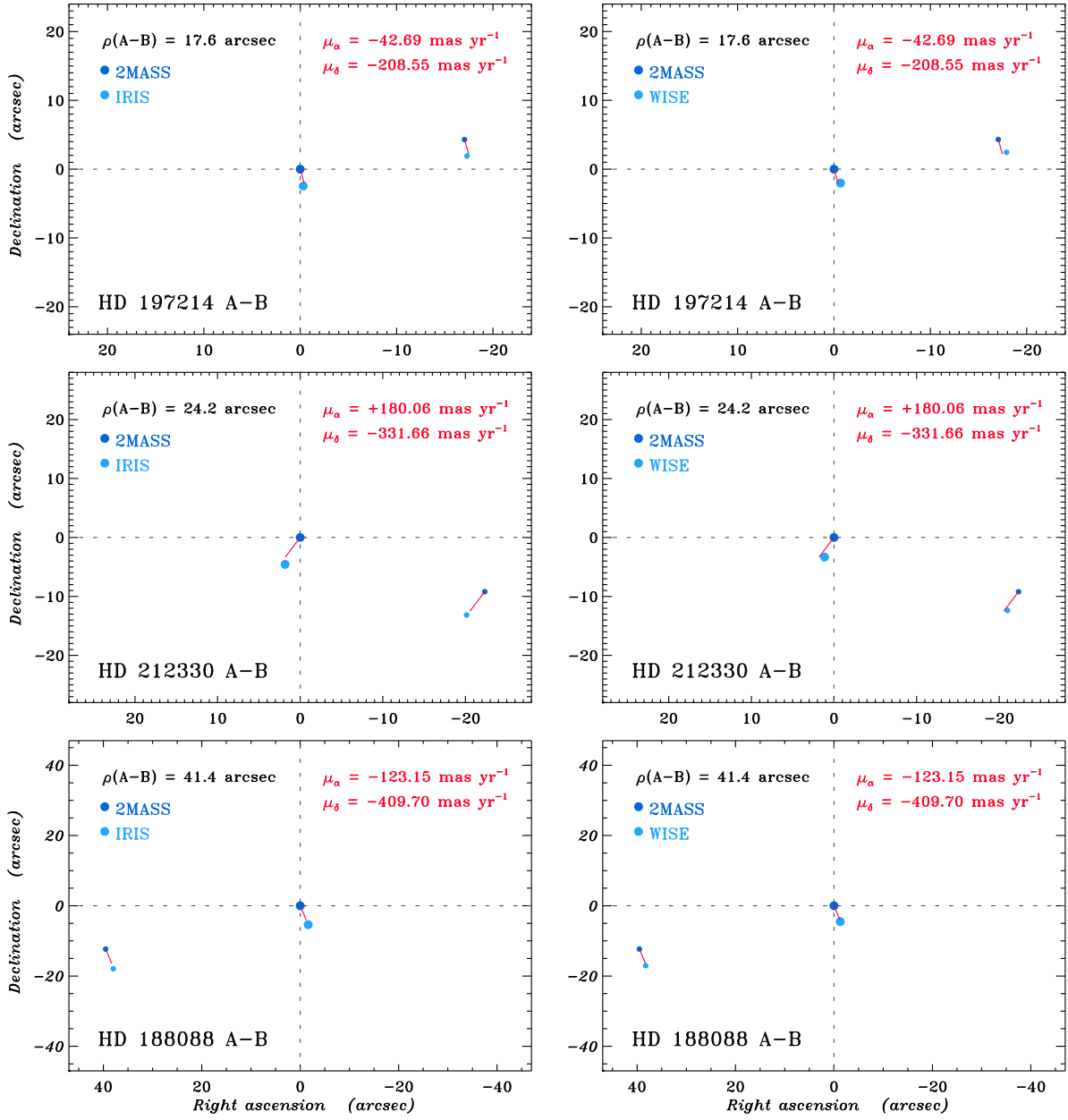}}
\caption{Same as Fig.~\ref{FIG_CPMS_1}, but for HD\,197214~A-B, HD\,212330~A-B, and HD\,188088~A-B.}
\label{FIG_CPMS_3}
\end{figure*}

\bsp

\label{lastpage}

\end{document}